\documentclass[aps,prl,twocolumn,showpacs,groupedaddress,psfig]{revtex4}

\usepackage{epsfig}
\usepackage{amssymb}

\newcommand{\beq}{\begin{equation}}
\newcommand{\bea}{\begin{eqnarray}}
\newcommand{\eeq}{\end{equation}}
\newcommand{\eea}{\end{eqnarray}}
\newcommand{\nn}{\nonumber}

\newcommand{\bx}{{\mathbf x}}

\begin{document}

\title{Simulating nonequilibrium quantum fields with stochastic 
quantization techniques}

\author{J.\ Berges$^1$}
\author{I.-O.\ Stamatescu$^{1,2}$}
\affiliation{$^1$Institute for Theoretical Physics, University of Heidelberg,\\
Philosophenweg 16, 69120 Heidelberg, Germany\\
$^2$FEST, Schmeilweg 5, 69118 Heidelberg, Germany}

\begin{abstract}
We present lattice simulations of nonequilibrium quantum fields 
in Minkowskian space-time. Starting from a non-thermal initial state, 
the real-time quantum ensemble in 3+1 dimensions 
is constructed by a stochastic process in an additional (5th) 
``Langevin-time''. For the example of a self-interacting scalar field 
we show how to resolve apparent unstable Langevin dynamics,
and compare our quantum results with those obtained in classical
field theory. Such a direct simulation method 
is crucial for our understanding of collision experiments of 
heavy nuclei or other nonequilibrium phenomena in strongly coupled
quantum many-body systems. 
\end{abstract}

\maketitle

Nonequilibrium quantum field theory is the tool
to understand a large variety of topical phenomena in
high-energy particle physics, cosmology as well as condensed 
matter physics. Current and future collision experiments of 
heavy nuclei involve far-from-equilibrium dynamics
for strongly interacting matter described by quantum chromodynamics 
(QCD). Other experiments, which have attracted much interest recently, 
concern the dynamics of ultra-cold quantum gases. 
Though these involve length scales many orders of 
magnitude larger than QCD, they require similar quantum field
theoretical techniques. 

For out-of-equilibrium calculations standard
approximation techniques, such as perturbation theory, 
are not uniform in time and fail to describe thermalization.   
There has been substantial progress in our analytical
understanding of nonequilibrium quantum fields using $n$-particle 
irreducible functional integral techniques~\cite{Berges:2004yj}.
However, nonequilibrium truncations are difficult 
to test for crucial questions of QCD or near a Feshbach 
resonance in atomic media, i.e.\ where strong interactions play an 
important role. Direct simulations on a space-time lattice could 
boost our knowledge and trigger the development of further approximate 
analytical tools. 

Despite the importance of non-perturbative lattice simulation techniques in 
out-of-equilibrium quantum field theory, these have not been developed 
so far. This is in sharp contrast to well-established thermal equilibrium
methods~\cite{lattice}. 
Equilibrium calculations can typically be based on 
a Euclidean formulation, where the time variable is analytically 
continued to imaginary values. By this the quantum theory is mapped 
onto a statistical mechanics problem, which can be simulated by 
importance sampling techniques. Nonequilibrium problems, 
however, are not amenable to a Euclidean formulation. 
Moreover, for real times standard importance sampling is not possible 
because of a non-positive definite probability measure.
Efforts to circumvent this problem include considering 
the computer-time evolution in Euclidean lattice 
simulations~\cite{computertime,Miller:2000pd}.
A problem in this case is to calibrate the computer time independently
of the algorithm.

In this Letter we present a nonequilibrium quantum field theory
simulation in Minkowskian space time. The quantum ensemble is 
constructed by a stochastic process in an additional ``Langevin-time'' 
using the reformulation of stochastic 
quantization~\cite{stochquant,Seiler:1983mz} for the Minkowskian path 
integral~\cite{cl,Minkowski}: The quantum fields are 
defined on a 3+1 dimensional physical
space-time lattice, while the updating procedure employs a Langevin 
equation with a complex driving force in a 5th, unphysical ``time'' 
direction. Nonequilibrium dynamics is implemented by
specifying an initial state or density matrix, 
which deviates from thermal equilibrium. For the example of
a relativistic scalar field theory with quartic self-interaction,
we compute the time evolution of
correlation functions and consider the 
characteristic damping rates. 

Though more or less formal proofs 
of equivalence of the stochastic approach and the path integral 
formulation have been given for Minkowski space-time, 
not much is known about the 
general convergence properties and its 
reliability beyond free-field theory or simple examples
in equilibrium~\cite{Minkowski}. 
Much more advanced applications   
concern simulations in Euclidean space-time with non-real 
actions~\cite{Ambjorn:1986mf,Karsch:1985cb}, where standard
Monte Carlo methods do not work. Despite successful examples, 
major reported problems concern unstable dynamics and 
incidences of apparent convergence to unphysical 
results~\cite{convergence,Ambjorn:1986mf,Karsch:1985cb,disc}. 

To our knowledge the approach has not been used to simulate 
nonequilibrium quantum field theory before, though some 
properties seem to make it quite suitable for that. Firstly,  
nonequilibrium requires specification of an initial state
or density matrix. Therefore, the initial configuration is
fixed which seems to stabilize the procedure. Moreover,
the additional averaging over an
initial density matrix can help to achieve fast convergence.
Secondly, one typically has a good guess for the $3+1$ dimensional starting 
configurations of the Langevin updating procedure:
In contrast to the quantum theory, the corresponding classical 
statistical field theory can be simulated using numerical 
integration and Monte Carlo 
techniques~\cite{Berges:2004yj}.
Using the nonequilibrium classical statistical solution as the 
starting configuration can improve convergence. It also provides 
a crucial check of the quantum result in some limiting 
cases: For sufficiently large macroscopic field or 
occupation numbers classical dynamics can 
provide a good approximation~\cite{Berges:2004yj}. 
 
For our example we observe good convergence properties
of the quantum simulations, which is a remarkable result. For given initial 
field configurations at time $t=0$, very different starting 
configurations for the 3+1 dimensional space-time lattice converge 
to the same nonequilibrium dynamics for all $t > 0$. 
To obtain this we had to resolve the 
problem of possible unstable dynamics for the updating procedure,
as is described in detail below.
We compare our quantum results with those obtained for the
corresponding classical theory for same initial 
conditions and lattice regularization.
We indeed find agreement in those cases where this is expected,
and observe increasing deviations for smaller fields
or occupation numbers. In the following we describe the
relevant theoretical ingredients and
present the numerical evidence.

{\em Nonequilibrium quantum field theory} can be described    
by the generating functional for correlation functions~\cite{Berges:2004yj}:
\bea
\lefteqn{Z[J;\rho] = 
{\rm Tr}\left\{ \rho\, T_{\mathcal C}\, e^{i \int_{\mathcal C}\!
J(x) \Phi(x)}\right\}} \nonumber\\
&=& \int {\rm d} \varphi_1 {\rm d} \varphi_2 \, 
\rho(\varphi_1,\varphi_2) 
\int\limits_{\varphi_1}^{\varphi_2}
[{\rm d} \varphi]\, e^{i \int_{\mathcal C} \left( L(x) 
+ J(x)\varphi(x)\right)} 
\label{eq:definingZneq}  .\,\,
\eea
The path integral (\ref{eq:definingZneq})
displays the quantum fluctuations for a theory with 
Lagrangian $L$, 
and the statistical fluctuations encoded 
in the weighted average with the initial-time density matrix
$\rho(\varphi_1,\varphi_2)$. 
Here $T_{\mathcal C}$ denotes contour time ordering along a 
closed path $\mathcal C$ starting at $t \equiv x^0 =0$ 
with \mbox{$\int_{\mathcal C} \equiv \int_{\mathcal C} 
{\rm d} x^0 \int {\rm d}^d x$} (usual 
time ordering along the forward piece $\mathcal C^+$, and anti-temporal 
ordering on the backward piece $\mathcal C^-$). 
The initial fields
are fixed by $\varphi_1(\bx) = \varphi(0^+,\bx)$ and
$\varphi_2(\bx) = \varphi(0^-,\bx)$.  
Nonequilibrium correlation functions, 
i.e.~expectation values of Heisenberg field operators $\Phi(x)$, are obtained
by functional differentiation. The two-point
function, e.g., is 
\bea
{\rm Tr}\{\rho\, T \Phi(x)\Phi(y)\} 
= \frac{\delta^2 Z[J;\rho]}
{i \delta J(x) i \delta J(y)}\Big|_{J=0}   
\label{eq:twopointneq}
\eea
with all time arguments on $\mathcal C^+$ such that $T_{\mathcal C}$ 
corresponds to standard time ordering $T$. In the following 
we consider physical correlation functions, which have their 
arguments on $\mathcal C^+$. The role of $\mathcal C^-$ is then 
only to normalize \mbox{$Z[J=0;\rho] = 1$} with ${\rm Tr} \rho = 1$.
\begin{figure}[t]
\vspace{4.9cm} 
\includegraphics{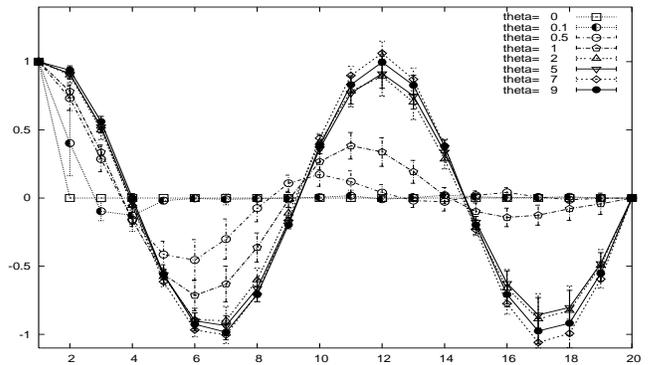} 
\caption{${\rm Re} C({\hat t})$ vs ${\hat t}$ for a free field 
theory with mass ${\hat m} =2.315$. The Langevin evolution,
shown for $\vartheta= 0$--$9$ in units of $a^2$, 
converges to the correct result
with period $2\pi \gamma/{\hat m}$.} 
\label{fig:free} 
\end{figure}

{\em Complex Langevin:}
The complex exponential weight in
(\ref{eq:definingZneq}) requires a simulation technique, which is not based 
on a probability interpretation. Stochastic quantization reformulated for 
real times~\cite{cl,Minkowski} can provide such an approach.
The stochastic process is described by a Langevin-type equation, which 
for a real quantum field theory governs a {\em complex} field 
$\phi = \phi_R + i \phi_I$.  
The appearance of an imaginary part reflects the
fact that in the quantum theory the field  
picks up a phase by evolving in time.
In addition to the space-time variable $x$ the field
depends on the Langevin-time parameter~$\vartheta$ with~\cite{cl,Nelson}
\bea
\frac{\partial \phi(x;\vartheta)}{\partial \vartheta} 
&=& i \frac{\delta S[\varphi]}{\delta \varphi(x)}\Big|_{\varphi \to \phi}
+ \eta(x;\vartheta)  \, .
\label{eq:RI}
\eea
Here $\delta S/\delta \varphi|_{\varphi \to \phi} = - \Box \phi
- m^2 \phi - \lambda \phi^3$
for a scalar theory with mass $m$ and  
self-interaction $\lambda$.
In general the real and imaginary part of the Gaussian noise 
term $\eta = \eta_R + i\eta_I$ 
can be both non-vanishing~\cite{Minkowski}, 
and the different choices 
may be used for optimizing convergence. 
We consider  $\eta_I \equiv 0$, with 
$\langle \eta(x;\vartheta) \rangle_\eta = 0$ and
\beq
\langle \eta(x;\vartheta)\eta(x';\vartheta') \rangle_\eta 
= 2 \delta(x-x') 
\delta(\vartheta-\vartheta') \, ,
\label{eq:noise}
\eeq
where $\langle \dots \rangle_\eta$ indicates average over the noise.

The stochastic process (\ref{eq:RI}) 
is associated to a distribution $P(\phi_R,\phi_I;\vartheta)$ and
averages of observables $A(\phi)$ are given as area integrals 
in the complex field plane:
\bea
\langle A \rangle_\eta &=& \frac{\int [{\rm d} \phi_R][{\rm d} \phi_I]
A(\phi_R + i \phi_I) P(\phi_R,\phi_I;\vartheta)}{\int [{\rm d} 
\phi_R][{\rm d} \phi_I] P(\phi_R,\phi_I;\vartheta)}
\nonumber\\
&=& \frac{\int [{\rm d} \phi_R]
A(\phi_R) P_{\rm eff}(\phi_R;\vartheta)}{\int [{\rm d} 
\phi_R] P_{\rm eff} (\phi_R;\vartheta)}\, . 
\label{eq:obs}
\eea
Here $P_{\rm eff}(\phi_R;\vartheta) \equiv \int [{\rm d} \phi_I]
P(\phi_R - i \phi_I,\phi_I;\vartheta)$,
where the shift in the integration variable 
$\phi_R \to \phi_R - i \phi_I$ for the second 
equality in (\ref{eq:obs}) is assumed to hold. The complex pseudo-distribution 
$P_{\rm eff}(\phi_R;\vartheta)$ is indeed governed
by the analytic continuation of the Fokker-Planck equation
to real times, which admits
the stationary solution~\cite{cl,Minkowski}
\bea
\lim_{\vartheta \to \infty} P_{\rm eff}(\phi_R;\vartheta) \equiv
P_{\rm eff} [\phi_R] \sim e^{i S[\phi_R] } \, . 
\eea
Thus the approach can in principle be used for a 
Minkowskian theory such as (\ref{eq:definingZneq}),
with ``ensemble'' averages calculated as averages
along Langevin trajectories. 

\begin{figure}[t]
\vspace{4.9cm} 
\includegraphics{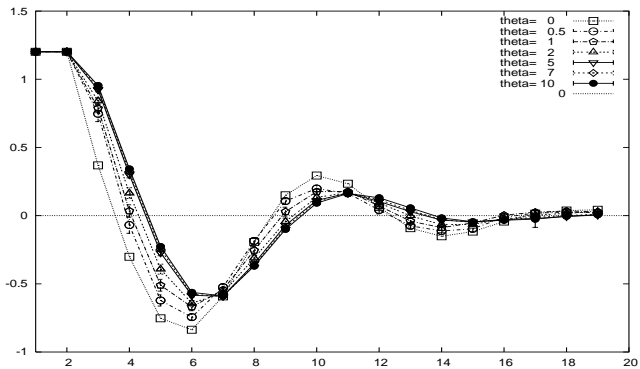} 
\caption{${\rm Re} G({\hat t})$ vs ${\hat t}$ for the 
interacting theory with $\lambda=1$. As  
 starting configuration ($\vartheta =0$)  
the classical result is taken, and the Langevin updating
incorporates quantum corrections.} 
\label{fig:int} 
\end{figure}
{\em Numerical simulation:} 
We consider $N_s^3 N_t$ lattices with 
anisotropic space-time discretization $a$ and $a_t$. 
Because of the Courant condition,
stable dynamics requires $a / a_t \equiv \gamma > \sqrt{d}$.
The Langevin-time discretization is $\delta\vartheta$. 
In terms of lattice variables
${\hat \phi} = a \phi$, ${\hat m} = a m$,
${\hat {\bf x}} = {\bf x}/a$, ${\hat t} = t/a_t$,  
${\hat \vartheta} = \vartheta/a^2$,  
$\epsilon = \delta\vartheta/a^2$, 
${\hat \eta} = \sqrt{a^3 a_t \delta\vartheta}\, \eta
= \sqrt{\epsilon/\gamma}\, a^3 \eta$,  
\beq
\langle{\hat \eta}({\hat x}, {\hat \vartheta} )\, 
{\hat \eta} ({\hat x}',{\hat \vartheta}')\rangle_\eta =
2\, \delta_{{\hat x},{\hat x}'}\delta_{{\hat \vartheta},{\hat \vartheta}'}
\eeq
the discretized equation (\ref{eq:RI}) in It{\^ o} calculus reads
\bea 
\lefteqn{
{\hat \phi}({\hat x};{\hat \vartheta} + \epsilon) 
= {\hat \phi}({\hat x};{\hat \vartheta}) 
+ \sqrt{\epsilon \gamma}\, {\hat \eta}({\hat x};{\hat \vartheta}) } 
\nonumber \\ 
&&- i \, \epsilon \,\left( \Box_{\gamma}
{\hat \phi}({\hat x};{\hat \vartheta})
+{\hat m}^2 {\hat \phi}({\hat x};{\hat \vartheta})
+\lambda {\hat \phi}({\hat x};{\hat \vartheta})^3 \right).
\quad  
\eea
Here $\Box_{\gamma}$ is the (anisotropic) lattice 
d'Alembertian \footnote{The updating proceeds sequentially 
in time and for bushes of independent variables in each time 
sheet. This improves convergence.}:  
\bea
\!\!\!\!\!\!&&\Box_{\gamma}{\hat \phi}({\hat x}) 
= \gamma^2 \left({\hat \phi}({\hat x}+{\hat e}_0) +
{\hat \phi}({\hat x}-{\hat e}_0) -c_t 
{\hat \phi}({\hat x}) 
\right) \nn \\
\!\!\!\!\!\!&&-\sum_i \left({\hat \phi}({\hat x}+{\hat e}_i) +
{\hat \phi}({\hat x}-{\hat e}_i) 
- 2 {\hat \phi}({\hat x}) \right) 
\eea
with $c_t=2$ for $1 < {\hat t} < N_t-1$ and $c_{N_t-1}=1$ for free 
large-${\hat t}$
boundary conditions (no coupling to ${\hat t}=N_t$). 
In this case we consider ${\hat \phi}({\hat t}=1,{\hat {\bf x}})
= {\hat \phi}({\hat t}=2,{\hat {\bf x}}) 
= {\hat \phi}_{\rm class}({\hat t}=1,{\hat {\bf x}})$
to set the {\em initial conditions}. Below we will also 
use $c_{N_t-1}=2$ for fixed large-${\hat t}$ b.c.\ 
in the case of a non-interacting 
field for comparison, and we set ${\hat \phi}({\hat t}=1,{\hat {\bf x}}) 
= 1$ 
and ${\hat \phi}({\hat t}=N_t,{\hat {\bf x}})=0$.
The classical field configurations
${\hat \phi}_{\rm class}({\hat t},{\hat {\bf x}})$ 
have been obtained by numerically solving the 
classical field equations and sampling over initial conditions,
with nonzero field average and Gaussian fluctuations~\cite{Berges:2004yj}.
Spatial p.b.c.\ are used. 

Here we speak of ``initial" configuration 
referring to the physical time, and of ``starting" configuration
for the Langevin process. As {\it starting} configurations the 
classical solution, i.e.\
${\hat \phi}({\hat t}>1,{\hat {\bf x}}; {\hat \vartheta}=0) =
 {\hat \phi}_{\rm class}({\hat t}>1,{\bf x}) $, or the
``null'' configuration ${\hat \phi}({\hat t}>1,{\bf x}; 
{\hat \vartheta}=0) \equiv 0$ are employed. 
The figures are for a 
$8^3 20$ lattice with $\gamma=4$ based on
$10^6$ updatings with $\epsilon = 10^{-5}$ ($\sim 2$ 
hours vector processor time). 
Error bars are statistical and only indicative.

In the following we present results for the two-point function 
(\ref{eq:twopointneq}). In Fig.~\ref{fig:free}, the 
correlator
\beq
C({\hat t}) = \langle \frac{1}{N_s^3} \sum_{\hat {\bf x}} 
{\hat \phi}(1, {\hat {\bf x}}) 
{\hat \phi}({\hat t}, {\hat {\bf x}})\rangle 
\label{eq:corr}
\eeq
for a {\em free} field of mass ${\hat m}=2.315$ is shown for
fixed b.c.\ with zero momentum initial configuration. $\langle \ \ldots
\rangle$ denotes average along the Langevin trajectories~\footnote{We note that 
${\hat \phi}(1, {\hat {\bf x}}) $ is not updated, therefore it can be taken out
of the brackets.}.
Shown are snapshots of ${\rm Re} C({\hat t})$ for 
Langevin-time parameter ${\hat \vartheta} = 0$--$9$, 
with null start configuration. 
The evolution in ${\hat \vartheta}$
exhibits slowly damped oscillations~\footnote{These oscillations do not
appear in the interacting case.}, converging to the 
free-field result with the correct ${\hat t}$-period of $2\pi\gamma/{\hat m}$.

\begin{figure}[t]
\vspace{4.9cm} 
\includegraphics{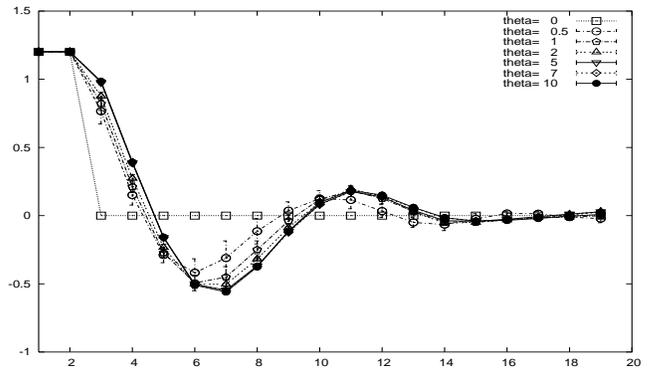} 
\caption{Same nonequilibrium initial condition as in Fig.~\ref{fig:int}, 
but null starting configuration for the Langevin updating. 
Comparison demonstrates the start independence of the result.} 
\label{fig:int1} 
\end{figure}
\begin{figure}[t]
\vspace{4.9cm} 
\includegraphics{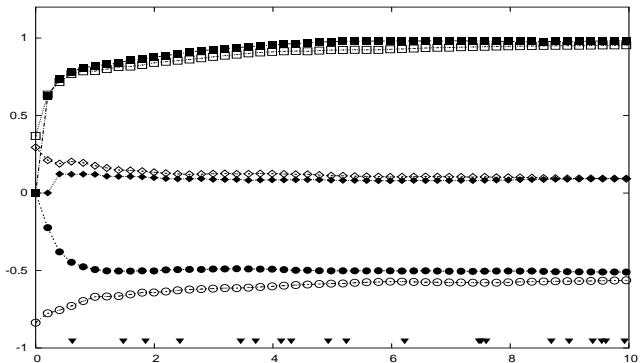} 
\caption{${\rm Re}G({\hat t})$ vs $\vartheta/a^2$ at
${\hat t} = $  3 (squares), 6(circles) and 10 (diamonds), for classical 
(open symbols) and null (full symbols) starting configurations. 
Initial conditions as in
Fig.~\ref{fig:int}.} 
\label{fig:convergence} 
\end{figure}

The unequal-time correlator (\ref{eq:corr})
measures the correlation of the field at time ${\hat t}$ with the 
initial field. It gives important information about the characteristic 
time scale for the loss of details about the initial conditions. In 
contrast to the free-field behavior, the interacting theory
has a finite characteristic damping time.
This is demonstrated in Fig.~\ref{fig:int},
which shows the connected part 
\beq
G({\hat t}) = C({\hat t})  
- \langle \frac{1}{N_s^3}\sum_{\hat {\bf x}}{\hat \phi}
(1, {\hat {\bf x}})\rangle 
\langle\frac{1}{N_s^3}\sum_{{\hat {\bf x}}'} {\hat \phi}({\hat t}, 
{\hat {\bf x}}')\rangle \,  
\label{eq:connected}
\eeq
for $\lambda=1$ and ${\hat m} = 0$. 
In Fig.~\ref{fig:int1} a different starting configuration
is considered for same ${\hat \phi}_{\rm class}(1,{\hat {\bf x}})$
initial condition as in Fig.~\ref{fig:int}. 
The same data is presented as a function of the Langevin-time 
${\hat {\vartheta}}$ in Fig.~\ref{fig:convergence} to see the convergence. 
For these parameters one expects moderate quantum effects.
In runs with larger coupling or smaller field we find that
the Langevin updating incorporates increasing quantum corrections.
Accordingly, one observes larger deviations compared
to the classical starting configuration.

In these simulations with $\epsilon =10^{-5}$ we encounter
incidences of unstable Langevin dynamics (see also \cite{convergence}).
Their appearance depends on the random number and they
are strongly suppressed by using a smaller step size, which indicates
that they are artefacts of the discretization. To cope with them we 
used two methods: 1) back-stepping on the trajectory 
some thousands steps (about 0.1 in ${\hat \vartheta}$) 
and restart with a new random number, and 2) regularizing the process
by a small imaginary mass (about $10^{-4}/a$) in the action. 
Both methods worked quite well. 
In Fig. 3 triangles indicate the back steppings with 
regularization for the null start. For $\epsilon =10^{-6}$ we
practically could eliminate the runaway trajectories on runs of 
the same ${\hat \vartheta}$ length
(but ten fold computer time).

Two procedures can be employed for further tests, which
are beyond the scope of this letter. Firstly, one can compare to
analytical approximations based on higher \mbox{$n$-particle} 
irreducible effective actions~\cite{Berges:2004yj}.     
Secondly, going to sufficiently late times 
one can compare to certain thermal equilibrium results from
Euclidean simulations. 

The numerics can be optimized by using improved space-time
derivatives and Langevin algorithms. A more systematic 
study of the convergence problems and
of the associated "Fokker-Planck" equation also has to be done.
This also includes volume, lattice discretization and step size 
dependence which lead to systematic 
effects~\cite{stochquant,Seiler:1983mz,cl,Minkowski,
Ambjorn:1986mf,Karsch:1985cb,convergence,disc}. 

We have demonstrated the possibility of first-principles
simulations in nonequilibrium quantum field dynamics.
The range of potential applications is enormous.
It may be used for out-of-equilibrium as well as 
Minkowskian equilibrium properties extracted at late times.
The scalar theory considered here extended to two components is 
already relevant for the dynamics of Bose condensates. 
Possible applications to QCD require implementation in a non-Abelian gauge 
theory, which is work in progress.

{\em Acknowledgments.} J.B.\ thanks the KITP, Santa Barbara and
I.O.S.\ thanks BNL, Brookhaven and AEI, Golm for hospitality 
and support while part of 
this work has been conducted. The calculations have been done 
on the VPP5000 computer of the University of Karlsruhe.

\end{document}